\documentclass[aps,prb,twocolumn,superscriptaddress,showpacs]{revtex4}
\usepackage{bm}
\usepackage{graphicx}
\usepackage{rotating}
\usepackage{epsfig}
\usepackage{amssymb}
\begin{document}
\def\bA{{\bf A}}
\def\ba{{\bf a}}
\def\d{{\bf d}}
\def\P{{\bf{K}}}
\def\bGamma{\bm \Gamma}
\def\bD{{\bf D}}
\def\bK{{\bf K}}
\def\bk{{\bf k}}
\def\bM{{\bf M}}
\def\bQ{{\bf Q}}
\def\bF{{\bf F}}
\def\bL{{\bf L}}
\def\bx{{\bf x}}
\def\bz{{\bf z}}
\def\bu{{\bf u}}
\def\bR{{\bf R}}
\def\br{{\bf r}}
\def\bq{{\bf q}}
\def\bp{{\bf p}}
\def\bG{{\bf G}}
\def\bGh{{\bf \hat{G}}}
\def\bs{{\bf s}}
\def\bj{\bf j}
\def\bV{{\bf V}}
\def\bv{{\bf v}}
\def\b0{{\bf 0}}
\def\la{\langle}
\def\ra{\rangle}
\def\bLam{\bm{\Lambda}}
\def\bnab{\bm{\nabla}}
\def\bfeta{\bm{\eta}}
\def\Im{\mathrm {Im}\;}
\def\Re{\mathrm {Re}\;}
\def\beq{\begin{equation}}
\def\eeq{\end{equation}}
\def\bea{\begin{eqnarray}}
\def\eea{\end{eqnarray}}
\def\bdm{\begin{displaymath}}
\def\edm{\end{displaymath}}
\def\bnab{\bm{\nabla}}
\def\bSig{\bm{\Sigma}}
\def\Tr{{\mathrm{Tr}}}
\def\bJ{{\bf J}}
\def\bU{{\bf U}}
\def\bPsi{{\bf \Psi}}
\def\bPhi{{\bf \Phi}}
\def\Dth{{\Delta^{(\mathrm{III})}_0}}
\def\Dtw{{\Delta^{(\mathrm{II})}_0}}
\def\mtw{{\mu^{(\mathrm{II})}}}
\def\mth{{\mu^{(\mathrm{III})}}}
\def\dblone{\hbox{$1\hskip -1.2pt\vrule depth 0pt height 1.6ex width 0.7pt
                  \vrule depth 0pt height 0.3pt width 0.12em$}}

\title{The Josephson relation for the superfluid density in the BCS-BEC crossover}
\author{Edward~Taylor}
\affiliation{CNR-INFM BEC Center and Dipartimento~di~Fisica, Universit\`a di Trento, I-38050 Povo, Trento, Italy}

\date{April 16, 2008}

\begin{abstract}
The Josephson relation for the superfluid density is derived for a Fermi superfluid in the BCS-BEC crossover.  This identity extends the original Josephson relation for Bose superfluids.  It gives a simple exact relation between the superfluid density $\rho_s$ and the broken-symmetry Cooper pair order parameter $\Delta_0$ in terms of the infrared limit of the pair fluctuation propagator.  The same expression holds through the entire BCS-BEC crossover, describing the superfluid density of a weak-coupling BCS superfluid as well as the superfluid density of a Bose condensate of dimer molecules. 
\end{abstract}


\maketitle
\section{Introduction}
Josephson's relation~\cite{Josephson,Baymbook} has played an important role in developing an understanding of superfluidity since it establishes the connection between the two order parameters that are widely used to discuss Bose superfluids: the condensate density $n_c$ and the superfluid density $\rho_s$.  It provides an exact relation between these quantities in a Bose superfluid in terms of the infrared behaviour of the single-particle Green's function $D_{11}$ for bosons of mass $m_B$:
\bea \rho_s = -\lim_{\bq\rightarrow 0}\frac{n_c m^2_B}{\bq^2D_{11}(\bq,0)}.\label{Josephson}\eea
Here, $D_{11}(\bq,0)$ is the single-particle Green's function for bosons with momentum $\bq$ at zero Matsubara frequency, $\nu_m=0$.  

The simple structure of Eq.~(\ref{Josephson}) has enabled a detailed analysis to be carried out of the superfluid transition in both three-~\cite{Josephson,Holzmann03} and two-dimensional~\cite{Holzmann05} Bose superfluids.  This includes finite-size systems where it has been used to study the Berezinskii-Kosterlitz-Thouless transition~\cite{KTB} in two-dimensional superfluids.  In particular, by starting from a finite two-dimensional system, Ref.~\onlinecite{Holzmann05} explicitly showed how a nonzero superfluid density can persist in the thermodynamic limit even though the condensate density vanishes.

Recently, Holzmann and Baym have extended the original phenomenological arguments of Josephson and given a microscopic proof of Eq.~(\ref{Josephson}) using diagrammatic perturbation theory.~\cite{Holzmann}  This proof extends earlier discussions by Bogoliubov,~\cite{Bog} Gavoret and Nozi\'eres,~\cite{GN} and Hohenberg and Martin~\cite{Hohmartin} at $T=0$ where $\rho_s = mn$.  A discussion of the Josephson relation for a Bose superfluid at finite temperatures is given by Griffin~\cite{Griffinbook} using the dielectric diagrammatic formalism (see also Wong and Gould~\cite{WG}).

Following Holzmann and Baym, in this paper the analogous exact relation for a two-component \textit{Fermi} superfluid is proven.  This gives the relationship between the superfluid density and the order parameter $\Delta_0$ that represents the Bose-condensate of Cooper pairs of fermions. 
The Josephson relation derived in this paper is analyzed in the BCS-BEC crossover~\cite{Eagles,Leggett,RanderiaGreenBook} picture of Fermi superfluids, widely studied in recent years in the context of ultracold atomic gases.~\cite{PhysRep,Trentoreview,Grimmreview,Dalibardreview}  In the BEC limit of this crossover, where the attractive interaction between the two species of fermions is strong, the Cooper pairs reduce to dimer molecules.  In this limit, the expression obtained for $\rho_s$ reduces to the usual Josephson relation for a Bose superfluid in Eq.~(\ref{Josephson}).

As with the derivations for a Bose superfluid in Refs.~\onlinecite{Holzmann,Bog,GN,Hohmartin,Griffinbook}, the derivation given below for a uniform system is based on exact two-fermion propagators, and is not approximate.

\section{Preliminaries}
Consider a two-component Fermi gas (e.g., neutral Fermi atoms prepared in two different hyperfine states) with $s$-wave interactions between the two components, described by the Hamiltonian density (in this paper, $\hbar$ and also the volume $V$ are set to unity)
\bea {\cal{H}} &=& \sum_{\sigma}\bar{\psi}_{\sigma}(x)\left(-\frac{\bnab^2}{2m_F}
-
\mu\right)\psi_{\sigma}(x) \nonumber\\&&- U_0\bar{\psi}_{\uparrow}(x)\bar{\psi}_{\downarrow}(x){\psi}
_{\downarrow}
(x){\psi}_{\uparrow}(x). \label{Hdef0} \eea
The two components are denoted by $\sigma = \uparrow,\downarrow$ and $m_F$ is the fermion mass.  The use of a momentum-independent pseudopotential interaction $U_0$ leads to ultraviolet divergencies that are regularized in the usual way by the Lippmann-Schwinger equation,
\bea \frac{1}{U_0} = -\frac{m_F}{4\pi a_s} + \sum_{\bk}\frac{m_F}{\bk^2}.\label{LS}\eea
Here, $a_s$ is the $s$-wave scattering length.  The entire BCS-BEC crossover can be probed by ``tuning" the $s$-wave scattering length from small and negative (BCS limit) through unitarity $(|a_s|=\infty)$, and finally into the BEC limit where $a_s$ is small and positive.

Although only the case of $s$-wave interactions between fermions is considered in this paper, the analysis given below can be extended to deal with a more general pairing interaction, as well as Hubbard-type Hamiltonians that describe fermions in a lattice.

The derivation of a Josephson relation for Fermi superfluids given in this paper makes use of the structure of the grand canonical thermodynamic potential $\Omega[v_s]$ of a current-carrying Fermi superfluid to identify the change in the free energy of a superfluid when a velocity is imposed on the condensate order parameter.  The superfluid density for a superfluid with velocity $v_s$ is then obtained from
\bea \rho_s= \left(\frac{\partial^2\Omega[v_s]}{\partial v^2_s}\right)_{\!\mu,\Delta_0}\left.\rule{0mm}{5.0mm}\right|_{v_s=0}.\label{rhosdef}\eea
It can be shown~\cite{TaylorPRA06} that this is equivalent to the standard definition~\cite{Fisher}
\bea \rho_s = \left.\frac{\partial^2 F[v_s]}{\partial^2 v^2_s}\right|_{v_s=0}\eea
given in terms of the free energy $F = \Omega + \mu n$.  One can also prove (see Appendix A in Ref.~\onlinecite{TaylorPRA06}) that Eq.~(\ref{rhosdef}) is equivalent to $\rho_s = \rho - \rho_n$, where $\rho=mn$ is the total mass density and the normal fluid density $\rho_n$ is given in terms of the transverse current correlation function.~\cite{Baymbook}  Explicitly,  for superfluid flow along the $z$-axis,
\bea \left(\frac{\partial^2\Omega[v_s]}{\partial v^2_s}\right)_{\!\mu,\Delta_0}\left.\rule{0mm}{5.0mm}\right|_{v_s=0} = \rho - m\langle\hat{J}_z\hat{J}_z\rangle,\label{JJ}\eea where $\hat{J}_z$ is the component of the total current operator in the $z$-direction.

Microscopically, the thermodynamic potential is given by the partition function ${\cal{Z}}$,
\bea \Omega = -\beta^{-1} \ln {\cal{Z}},\label{Omegadef}\eea
where $\beta \equiv (k_B T)^{-1}$.  Functional integral techniques allow us to express the partition function as a functional integral over fermionic Grassmann
fields $\psi$ and $\bar{\psi}$ as~\cite{Popovbook}
\bea {\cal{Z}} = \int {\cal D}[\psi, \bar{\psi}] e^{-S[\psi, \bar{\psi}]}.\label{Zdef}
\eea  The imaginary-time action
$S[\psi,\bar{\psi}]$ in Eq.~(\ref{Zdef}) for a two-component Fermi superfluid is given by 
\bea S[\psi, \bar{\psi}] = \int d^4x\;\left[\sum_{\sigma}\bar{\psi}_{\sigma}(x)\partial_{\tau}{\psi}_{\sigma}(x)
+ {\cal{H}}\right], \label{Sdef0} \eea where the Hamiltonian density ${\cal{H}}$ is given by Eq.~(\ref{Hdef0}).
Here, $x = ({\br, \tau})$ is used to denote the spatial coordinate $\br$ and the imaginary time $\tau = i t$, and $ \int d^4x \equiv \int_0^\beta d\tau \int d{\bf{r}}$.

A central aspect of the analysis in this paper (and that of Josephson) is the existence of a broken-symmetry order parameter in the superfluid phase.  For Fermi superfluids, this order parameter $\Delta_0$ is given by the anomalous average \bea \Delta_0 \equiv \frac{ U_0}{\beta}\sum_k\langle c_{\downarrow,-k}c_{\uparrow,k}\rangle.\label{gapdef}\eea
Here, $c_{\sigma,k}$ is the Fourier-transform of the Fermi Grassmann field $\psi_{\sigma}(x)$,
\bea \psi_{\sigma}(x) = \frac{1}{\sqrt{\beta}}\sum_k c_{\sigma,k}e^{ik\cdot x},\eea
where $k\equiv (\bk,\omega_n)$ is a 4-vector for the momentum $\bk$ and Fermi Matsubara frequency $\omega_n = \pi(2n+1)/\beta$, $n = 0,\pm 1,\pm 2,...$, and $k\cdot x \equiv \bk\cdot\br - \omega_n\tau$.  

In order to introduce the bosonic order parameter into the partition function ${\cal{Z}}$, the following identity is used:
\bea \lefteqn{e^{U_0\int d^4x\;\bar{\psi}_{\uparrow}
\bar{\psi}_{\downarrow}{\psi}_{\downarrow}{\psi}_{\uparrow}} =
\int
{\cal{D}}[\Delta,\Delta^*]\times}&&\nonumber\\
&&\!\!\!\!\!\!\!\exp\Big\{-\int d^4x\;\Big[\frac{|\Delta|^2}{U_0}-\left(\Delta^*{\psi}_{\downarrow}{\psi}_{\uparrow} +
\Delta\bar{\psi}_{\uparrow}\bar{\psi}_{\downarrow}\right)\Big]\Big\}.\label{HS}\eea  Substituting this into Eq.~(\ref{Zdef}), one obtains the result
\bea {\cal{Z}}&=&\int {\cal D}[\psi,
\bar{\psi}]{\cal D}[\Delta,\Delta^*]\times\nonumber\\&&\exp\Big\{-\int d^4x\Big[ \sum_{\sigma}\bar{\psi}_{\sigma}(x)\left(\partial_{\tau}
- \frac{\bnab^2}{2m} - \mu\right){\psi}_{\sigma}(x) \nonumber\\&& -
\Delta^*{\psi}_{\downarrow}{\psi}_{\uparrow} -
\Delta\bar{\psi}_{\uparrow}\bar{\psi}_{\downarrow} +
\frac{|\Delta|^2}{U_0}\Big]\Big\}\nonumber\\
&\equiv&
\int {\cal D}[\psi,\bar{\psi}]{\cal D}[\Delta,\Delta^*]\;e^{-S_{\mathrm{eff}}}.
\label{Z2}\eea
This integral identity (a Hubbard-Stratonovich transformation~\cite{HS}) only introduces an auxiliary Bose field $\Delta(x)$, and no approximation has been made.  It is straightforward to show that the static, uniform component of this field gives the order parameter defined in Eq.~(\ref{gapdef}).  That is, using the partition function in Eq.~(\ref{Z2}), one can show that
\bea \langle \Delta(x)\rangle = \frac{ U_0}{\beta}\sum_k\langle c_{\downarrow,-k}c_{\uparrow,k}\rangle \equiv \Delta_0,\label{gapdef2}\eea
where the equilibrium average $\langle\Delta(x)\rangle$ is defined by \bea \langle\Delta(x)\rangle \equiv \frac{1}{{\cal{Z}}}\int{\cal{D}}[\psi,\bar{\psi}] {\cal{D}}[\Delta,\Delta^*]\Delta(x)e^{-S_{\mathrm{eff}}}.\label{avgD}\eea
It is important to emphasize that while the BCS order parameter $\Delta_0$ is usually calculated in the mean-field BCS approximation, it is not an inherently mean-field quantity.  The ``0" subscript on $\Delta_0$ only denotes the fact that the order parameter is related to the average occupation (macroscopic in the superfluid phase) of a pair state with zero total momentum.  One obtains the mean-field approximation for $\Delta_0$ if the expectation value $\langle\cdots\rangle$ in Eq.~(\ref{gapdef2}) is evaluated using a mean-field expression for the partition function.  Here the full partition function is used, so $\Delta_0$ is the \textit{exact} value of the order parameter.

Having established the relation between the auxiliary Bose field $\Delta(x)$ and the order parameter in Eq.~(\ref{gapdef2}), $\Delta(x)$ can be separated as
\bea \Delta(x) = \Delta_0 + \Lambda(x),\label{Bogshift}\eea
where $\Lambda(x)$ represents the fluctuations out of the static Bose-condensed pair state.

The partition function given by Eq.~(\ref{Z2}) and the identity in Eq.~(\ref{Bogshift}) will be used below to analyze the superfluid density in a current-carrying Fermi superfluid.  First we examine the \textit{pair fluctuation propagator} that describes the dynamics of the Bose field $\Delta(x)$.

\section{The pair fluctuation propagator}
\label{secCP}
The key correlation function of interest in the study of the dynamics of Cooper pairs is the $2\times2$ matrix pair fluctuation propagator $\bL(x,x')$ that describes the propagation of the Bose field $\Delta(x)$.~\cite{LarkinVarlamov}  It is conveniently defined in terms of its inverse $\bL^{-1}$, with matrix elements given by
\bea (\bL^{-1})_{11}(x,x') &\equiv& -\frac{1}{U_0}\delta(x-x') + K_{11}(x,x')\nonumber\\&=&(\bL^{-1})_{22}(x',x)
\label{L11}\eea
and
\bea (\bL^{-1})_{12}(x,x') &\equiv& K_{12}(x,x'), \nonumber\\ (\bL^{-1})_{21}(x,x') &\equiv& K_{21}(x,x'). \label{L12}\eea
Here the matrix two-particle Green's function $\P$ is defined by~\cite{note}
\bea \P(x,x')\equiv\left [
\begin{array}{cc} \langle \bar{\Phi}(x)\Phi(x')\rangle  & \langle \Phi(x)\Phi(x')\rangle 
\\ \langle \bar{\Phi}(x)\bar{\Phi}(x')\rangle &
\langle \Phi(x)\bar{\Phi}(x')\rangle
\end{array}  \right],  \label{F} \eea
where \bea \bar{\Phi}(x)\equiv \bar{\psi}_{\uparrow}(x)\bar{\psi}_{\downarrow}(x), \;\;\Phi(x)\equiv \psi_{\downarrow}(x)\psi_{\uparrow}(x).\label{Phi}\eea  
Using Eqs.~(\ref{L11}) and (\ref{L12}), the pair fluctuation propagator is given explicitly by 
\bea \bL &\equiv&- \frac{U_0}{(1-U_0K_{11})(1-U_0K_{22}) - U^2_0K_{12}K_{21}}\times\nonumber\\&&\left (
\begin{array}{cc} 1 - U_0K_{22} & U_0K_{12}

\\
U_0K_{21}
&1-U_0K_{11}
\end{array} \right).\label{L}
\eea

The pair fluctuation propagator can be viewed as the propagator for a single composite boson (the Cooper pair) and hence, as an analogue of the single-boson Green's function $\bD$.   It should be stressed, however, that it is distinct from the two-particle Green's function $\P$, as can be seen from Eq.~(\ref{L}).  Nevertheless, as discussed in Ref.~\onlinecite{Pieri03}, evaluating the pair fluctuation propagator at the mean-field BCS level (equal to the BCS approximation for the many-body $T$-matrix in Ref.~\onlinecite{Pieri03}), one can show that $\bL$ reduces (up to a constant) to the single-particle Bose Green's function [within the Bogoliubov approximation; see Eq.~(\ref{Dtilde})] for a dimer condensate in the BEC limit of the BCS-BEC crossover.  From this point of view, $\bL$ is seen to be the natural analogue in a Fermi superfluid of the Bose Green's function $\bD$.  Further discussion of the relation between the pair fluctuation propagator and the two-particle Green's function $\P$ can be found in Ref.~\onlinecite{LarkinVarlamov}.

Motivated by the similarity to the Bose Green's function $\bD$, we expand the Fourier transform $\bL^{-1}(\bq,\nu_m)$ of the inverse pair fluctuation propagator in powers of $q$ as~\cite{note2}
\bea (\bL^{-1})_{11}(\bq,\nu_m) &=& (\bL^{-1})_{22}(\bq,-\nu_m) = Ai\nu_m - B -\nonumber\\&& C\bq^2 + {\cal{O}}(\bq^4,\nu^2_m),\nonumber
\\
(\bL^{-1})_{12}(\bq,\nu_m) &=&  (\bL^{-1})_{21}(\bq,\nu_m) = -D  - F\bq^2 \nonumber\\&& + {\cal{O}}(\bq^4,\nu^2_m), \label{Mexp}\eea
where $\nu_m = 2\pi m/\beta$, $m = 0,\pm 1,\pm 2,...$ denotes the Bose Matsubara frequencies.  Note that the off-diagonal element $(\bL^{-1})_{12}$ has the symmetry  $(\bL^{-1})_{12}(\bq,-\nu_m) = (\bL^{-1})_{12}(\bq,\nu_m)$ and hence the absence of a linear term in the Matsubara frequency in its expansion.  

Using the expansion in Eq.~(\ref{Mexp}), the pair fluctuation propagator becomes
\bea \lefteqn{\bL(\bq,\nu_m) = \frac{1}{A(i\nu_m - \omega_{\bq})(i\nu_m + \omega_{\bq})}\times}&&\nonumber\\
&&\left (
\begin{array}{cc} i\nu_m + \frac{B}{A} +\frac{C}{A}\bq^2 &
-\frac{D}{A}-\frac{F}{A}\bq^2
\\
-\frac{D}{A}-\frac{F}{A}\bq^2
&-i\nu_m + \frac{B}{A} +\frac{C}{A}\bq^2
\end{array} \right) + {\cal{O}}(\bq^4,\nu^2_m),\nonumber\\ \label{Lexp}
\eea
where we have defined the poles of $\bL$ as
\bea \omega_{\bq} \equiv \frac{1}{A}\sqrt{(B\!+\!C\bq^2)^2 \!-\! (D\!+\!F\bq^2)^2}. \eea
This shows that in order for the poles of the pair fluctuation propagator to be gapless, we must have $B=D$, giving a Bogoliubov-Anderson mode with velocity
\bea v =   \frac{1}{A}\sqrt{2B(C-F)}.\label{BA}\eea

The condition for a gapless mode to exist can also be written as
\bea \lefteqn{(\bL^{-1})_{11}(0,0) + (\bL^{-1})_{22}(0,0)-}&&\nonumber\\&&(\bL^{-1})_{12}(0,0) - (\bL^{-1})_{21}(0,0) = 0,\label{gapless}\eea
a result that will be made use of in deriving the Josephson relation in Sec.~\ref{secJR}.

\section{The current-carrying superfluid}
\label{secJR}
We now consider the properties of a current-carrying superfluid.  
To introduce a finite superfluid velocity $v_s$, a ``phase twist"~\cite{Fisher} is applied to the Bose order parameter $\Delta_0$ defined in Eq.~(\ref{gapdef}):
\bea \Delta_0\rightarrow \Delta_0e^{im_B\bv_s\cdot\br}.\label{pt}\eea  Here, $m_B \equiv 2m_F$ is the mass of the Cooper pair.  Expanding the Bose pairing field $\Delta(x)$ about the uniform, static value of the order parameter $\Delta_0$ as in Eq.~(\ref{Bogshift})
and applying the phase twist, one finds 
\bea \Delta(x) \!&\rightarrow&\! \Delta_0e^{im_B\bv_s\cdot\br} \!+\! \Lambda(x)
=\Delta(x)\! +\! \Delta_0\left(e^{im_B \bv_s \cdot \br}-1\right).\nonumber\\ \label{Dexp}\eea
Writing the dependence on the superfluid velocity $\bv_s$ in this way emphasizes that the proceeding analysis is based on the full Bose pairing field $\Delta(x)$ that includes all fluctuations about the static order parameter $\Delta_0$.  

Using Eq.~(\ref{Dexp}) in Eq.~(\ref{Z2}), it is seen that the effect of imposing a phase twist on the order parameter is to generate a new term in the effective action:
\bea S_{\mathrm{eff}}[v_s] = S_{\mathrm{eff}}[0] + \delta S[v_s],\label{Sexp}\eea
where
\bea \delta S &=& -\Delta_0  \int d^4x[(e^{im_B\bv_s\cdot\br}-1)\bar{\Phi}(x) \nonumber\\ &&
\;\;\;\;\;\;\;\;\;\;\;\;+ (e^{-im_B\bv_s\cdot\br}-1)\Phi(x)], \label{Sv1}\eea
and $\Phi,\bar{\Phi}$ are defined in Eq.~(\ref{Phi}). 
Note that the term in the effective action $S_{\mathrm{eff}}$ in Eq.~(\ref{Z2}) involving $|\Delta(x)|^2/U_0$ is unchanged by the phase twist to the order parameter since the imaginary-time integral over any term linear in $\Lambda(x)$ vanishes.

Using Eq.~(\ref{Sexp}) in the partition function defined in Eq.~(\ref{Z2}), the thermodynamic potential of the current-carrying superfluid is found to be
\bea \Omega[v_s] = -\frac{1}{\beta}\ln\int {\cal D}[\psi,\bar{\psi}]{\cal D}[\Delta,\Delta^*] e^{-S_{\mathrm{eff}}[0]-\delta S[v_s]}\label{Omegavs}.\eea
Applying the definition of the superfluid density given by Eq.~(\ref{rhosdef}) to Eq.~(\ref{Omegavs}), one finds
\bea \rho_s =  \frac{1}{\beta}\Big\langle \frac{\partial^2 \delta S}{\partial v^2_s}\Big\rangle- \frac{1}{\beta}\Big\langle\left(\frac{\partial \delta S}{\partial v_s}\right)^2\Big\rangle.\label{rhosdS}\eea
Here, $\langle\cdots\rangle$ denotes the equilibrium average in the current-free state, given by
\bea \langle\cdots\rangle \equiv \frac{1}{{\cal{Z[{\mathrm{0}}]}}}\int{\cal{D}}[\psi,\bar{\psi}] {\cal{D}}[\Delta,\Delta^*]\left(\cdots\right)e^{-S_{\mathrm{eff}}[0]}.\label{avg}\eea  Note that $\langle (\partial \delta S/\partial v_s)\rangle|_{v_s=0} = 0$, by symmetry.  

Using Eq.~(\ref{Sv1}) to evaluate Eq.~(\ref{rhosdS}) gives
\bea \lefteqn{\rho_s = \frac{\Delta_0m^2_B}{\beta}\int d^4x\; (\hat{\bv}_s\cdot\br)^2\langle\bar{\Phi}(x)+\Phi(x)\rangle}&&\nonumber\\&&\;\;\;\;-\frac{\Delta^2_0m^2_B}{\beta}\int d^4x\;d^4x'(\hat{\bv}_s\cdot\br)(\hat{\bv}_s\cdot\br')
[K_{11}+K_{22}\nonumber\\&&\;\;\;\;-K_{12}
-K_{21}](x,x'), \label{OmegaB}\eea
where $\hat{\bv}_s\equiv \bv_s/v_s$ and $K_{ij}(x,x')$ denote the elements of the matrix two-particle Green's function defined in Eq.~(\ref{F}). 
Making use of the identity
\bea \langle\bar{\Phi}(x)\rangle = \langle\Phi(x)\rangle = \frac{\Delta_0}{U_0},\eea
Eq.~(\ref{OmegaB}) is naturally given in terms of the inverse pair fluctuation propagator,
\bea \rho_s &=& -\frac{\Delta^2_0m^2_B}{\beta}\!\int\! d^4x\;d^4x'(\hat{\bv}_s\cdot\br)(\hat{\bv}_s\cdot\br')
[(\bL^{-1})_{11}\nonumber\\&&+(\bL^{-1})_{22}-(\bL^{-1})_{12}
-(\bL^{-1})_{21}](x,x'). \label{OmegaB3}\eea
Fourier transforming this expression, one finds (taking $\hat{\bv}_s = \hat{\bz}$ to lie along the $z$-axis)
\bea \rho_s &=& -\frac{\Delta^2_0m^2_B}{2}\! \lim_{\bq\rightarrow 0}\frac{\partial^2}{\partial q^2_z}[(\bL^{-1})_{11}+(\bL^{-1})_{22}-(\bL^{-1})_{12}\nonumber\\&&-(\bL^{-1})_{21}](\bq,0), \label{OmegaB2}\eea
with static matrix elements $(\bL^{-1})_{ij}(\bq,\nu_m=0)$.  In arriving at this result, a gapless Bose excitation spectrum has been assumed, using the result in Eq.~(\ref{gapless}).

Using the expansion in Eq.~(\ref{Mexp}) to evaluate the second-order derivative in Eq.~(\ref{OmegaB2}), we see that
\bea \lim_{\bq\rightarrow 0}\frac{\partial^2}{\partial q^2_z}[(\bL^{-1})_{11}\!+\!(\bL^{-1})_{22}\!-\!(\bL^{-1})_{12}\!-\!(\bL^{-1})_{21}](\bq,0)&&\nonumber\\&& \!\!\!\!\!\!\!\!\!\!\!\!\!\!\!\!\!\!\!\!\!\!\!\!\!\!\!\!\!\!\!\!\!\!\!\!\!\!\!\!\!\!\!\!\!\!\!\!\!\!\!\!\!\!\!\!\!\!\!\!\!\!\!\!\!\!\!\!\!\!\!\!=4F-4C.\label{key2}\eea
This allows us to more compactly write Eq.~(\ref{OmegaB2}) as
\bea \rho_s &=& 2\Delta^2_0m^2_B (C - F). \label{JF0}\eea
Now, from the static $(1,\!1)$ matrix element $L_{11}(\bq,0)$ in Eq.~(\ref{Lexp}), one also finds
\bea \lim_{\bq\rightarrow 0}\frac{1}{\bq^2L_{11}(\bq,0)} = \lim_{\bq\rightarrow 0}\frac{[D^2\!-\!B^2 \!+\! 2(DF\!-\!BC)\bq^2]}{B\bq^2}.\nonumber\\
\eea
Assuming that $\bL$ has a gapless excitation spectrum (such that $B=D$), this reduces to
\bea \lim_{\bq\rightarrow 0}\frac{1}{\bq^2L_{11}(\bq,0)} = 2F-2C.\label{key3}
\eea

Comparing Eqs.~(\ref{JF0}) and (\ref{key3}), one finally obtains
\bea \rho_s=-\lim_{\bq\rightarrow 0}\frac{\Delta^2_0m^2_B}{\bq^2L_{11}(\bq,0)}.\label{JF}\eea
This expression gives the precise analogue for Fermi superfluids of the Josephson relation for Bose superfluids in Eq.~(\ref{Josephson}).  We see that the single-particle Green's function $\mathbf{D}$ for bosons has been replaced by the pair fluctuation propagator $\bL$, and the square of the BCS order parameter $\Delta^2_0$ plays the role of the square of the order parameter \bea \Phi^2_0\equiv |\langle\psi\rangle|^2 = n_c \label{ncB}\eea
of a Bose superfluid.

For a Bose superfluid, Eq.~(\ref{ncB}) gives a simple relation between the order parameter $\Phi_0$ and the condensate density $n_c$, and these two quantities can be interchanged in the Josephson relation in Eq.~(\ref{Josephson}).  This is not the case in a Fermi superfluid, however, where the condensate density is not a simple function of the order parameter $\Delta_0$.  This can be seen from the mean-field expression for the condensate density in a BCS superfluid, given by~\cite{TaylorPRA07}
\bea n_c = \frac{1}{\beta^2}\sum_{\bk,\omega_n,\omega'_n}G_{0,21}(\bk,\omega_n)G_{0,12}(\bk,\omega'_n),\label{nc}\eea
where $\bG_0$ is the mean-field $2\times 2$ matrix BCS Green's function.  Equation~(\ref{JF}) emphasizes the direct role of the order parameter in Josephson's relation, in contrast to the indirect role played by the condensate density.

Equation~(\ref{JF}) gives an \textit{exact} relation between the superfluid density $\rho_s$ and the order parameter $\Delta_0$ in terms of the static pair fluctuation propagator $\bL(\bq,0)$.  It can immediately be used to study superfluidity in Fermi superfluids.  In Sec.~\ref{secapprox}, the Josephson relation is studied within the BCS approximation for the pair fluctuation propagator.  We see how the resulting expression for the superfluid density reduces to the Landau formula for BCS quasiparticle excitations.

\section{Relation to Landau's formula for a BCS superfluid}
\label{secapprox}
An important check of the Josephson relation for Fermi superfluids is that it reproduces Landau's well-known formula for the superfluid density when the normal fluid is comprised of BCS quasiparticle excitations,~\cite{statphys2}
\bea \rho_s = \rho + 2\sum_{\bk}\frac{\partial f}{\partial E_{\bk}} k^2_z.\label{Lrhos}\eea
Here, $f = [\exp(\beta E_{\bk}) +1]^{-1}$ is the Fermi thermal distribution for BCS quasiparticles of energy $E_{\bk} =  \sqrt{\xi^2_{\bk} + \Delta^2_0}$ and $k_z$ is the $z$-component of $\bk$.  The total mass density $\rho$ is given by\bea \rho = m_F\sum_{\bk}\left[1-\frac{\xi_{\bk}}{E_{\bk}}(1-2f)\right].\label{rhomf}\eea

The result given by Eqs.~(\ref{Lrhos}) and (\ref{rhomf}) for the superfluid density is mean-field insofar as it ignores the contribution to the normal fluid arising from bosonic collective modes, as discussed in Refs.~\onlinecite{Pieri03,TaylorPRA06}.  Consequently, we should be able to reproduce Eq.~(\ref{Lrhos}) by evaluating the Josephson relation in Eq.~(\ref{JF}) using a mean-field BCS approximation.  

Evaluating the pair fluctuation propagator $\bL^{-1}(\bq,0)$ within the BCS mean-field approximation amounts to evaluating the two-particle Green's function defined in Eq.~(\ref{F}) as a loop of two single-particle mean-field BCS Green's functions: $K = \sum G_{0}G_{0}$ (schematically).  Explicitly, Eqs.~(\ref{L11}) and (\ref{L12}) become
\bea \lefteqn{(\bL^{-1})_{11}\!(\bq,0) = (\bL^{-1})_{22}(\bq,0)=}&&\nonumber\\&&-\frac{1}{U_0}\! -\!  \frac{1}{\beta}\sum_{\bk,\omega_n}G_{0,11}(\bk,\omega_n)G_{0,22}(\bk-\bq,\omega_n)\label{F11mf}\eea
and \bea \lefteqn{(\bL^{-1})_{12}(\bq,0) = (\bL^{-1})_{21}(\bq,0)=}&&\nonumber\\&& -\frac{1}{\beta}\sum_{\bk,\omega_n} G_{0,12}(\bk,\omega_n)G_{0,12}(\bk-\bq,\omega_n). \label{F12mf} \eea

In this approximation, $\bL^{-1}$ is equivalent to the negative of the inverse pair fluctuation propagator $\bM$ defined in Refs.~\onlinecite{Engelbrecht, TaylorPRA06}.  Reference~\onlinecite{Engelbrecht} showed explicitly that the poles of $\bL$ describe the gapless (i.e., $B=D$) Bogoliubov-Anderson spectrum at small $\bq$ throughout the entire BCS-BEC crossover.  Furthermore, the combination of inverse matrix elements \bea (\bL^{-1})_{11}(\bq,0) - (\bL^{-1})_{12}(\bq,0) = -(C-F)\bq^2+\cdots\eea that enters the expression for the superfluid density in Eqs.~(\ref{OmegaB2}) and (\ref{JF0}) is proportional to the static inverse propagator for the \textit{phase} fluctuations of the order parameter.~\cite{note4}

Evaluating the second-order derivative of Eqs.~(\ref{F11mf}) and (\ref{F12mf}) with respect to $q_z$, after some lengthy but straightforward algebra, one finds
\bea C-F &=& \sum_{\bk}\frac{1}{4E^2_{\bk}}\left[\frac{1-2f}{2E_{\bk}} +  \frac{\partial f}{\partial E_{\bk}}\right]g(\bk)\nonumber\\&&+\sum_{\bk}\frac{\partial^2f}{\partial E^2_{\bk}}\frac{\xi^2}{4E^3_{\bk}}\left(\frac{k_z}{m_F}\right)^2, \label{BDmf}\eea
where $g(\bk)$ is defined by
\bea g(\bk) \equiv \frac{\xi_{\bk}}{m_F} - \left(\frac{k_z}{m_F}\right)^2\left[1 - 3\frac{\Delta^2_0}{E^2_{\bk}}\right].\label{g}\eea
Note that Eq.~(\ref{BDmf}) is the finite-temperature generalization of the $Q$ coefficient defined in Ref.~\onlinecite{Engelbrecht}.  

Substituting Eq.~(\ref{BDmf}) into Eqs.~(\ref{key3}) and (\ref{JF}) gives the following mean-field expression for the superfluid density:
\bea \rho_s &=& 2m^2_F\sum_{\bk}\frac{\Delta^2_0}{E^2_{\bk}}\left[\frac{1-2f}{2E_{\bk}} + \frac{\partial f}{\partial E_{\bk}}\right]g(\bk)\nonumber\\&&+2\sum_{\bk}\frac{\partial^2f}{\partial E^2_{\bk}}\frac{\xi^2\Delta^2_0}{E^3_{\bk}}k^2_z. \label{rhosmf}\eea
Using \bea \frac{\partial f}{\partial E_{\bk}} = \frac{m_FE_{\bk}}{k\xi_{\bk}}\frac{\partial f}{\partial k}\label{dervid}\eea and integrating by parts, Eq.~(\ref{rhosmf}) can be rewritten as
\bea \rho_s &=& 2m^2_F\sum_{\bk}\frac{\Delta^2_0}{2E^3_{\bk}}(1-2f)g(\bk)+2\sum_{\bk}\frac{\Delta^4_0}{E^4_{\bk}}\frac{\partial f}{\partial E_{\bk}}k^2_z\nonumber\\
&=&2\sum_{\bk}\frac{\Delta^2_0}{E^2_{\bk}}\left[\frac{1-2f}{2E_{\bk}} + \frac{\partial f}{\partial E_{\bk}}\right]k^2_z\nonumber\\&&+2m_F\sum_{\bk}\frac{\Delta^2_0}{2E^3_{\bk}}(1-2f)\left[\xi_{\bk} + \frac{k^2_z}{m_F}\left(\frac{3\Delta^2_0}{E^2_{\bk}} - 2\right)\right]\nonumber\\ &&-2\sum_{\bk}\frac{\Delta^2_0\xi^2_{\bk}}{E^4_{\bk}}\frac{\partial f}{\partial E_{\bk}}k^2_z. \label{rhosmf2}\eea  Applying Eq.~(\ref{dervid}) to the last line and integrating by parts again, one finds
\bea 
\rho_s &=&2\sum_{\bk}\frac{\Delta^2_0}{E^2_{\bk}}\left[\frac{1-2f}{2E_{\bk}} + \frac{\partial f}{\partial E_{\bk}}\right]k^2_z\nonumber\\&&+2m_F\sum_{\bk}\frac{\Delta^2_0}{2E^3_{\bk}}\left[\xi_{\bk} + \frac{k^2_z}{m_F}\left(\frac{3\Delta^2_0}{E^2_{\bk}} - 2\right)\right].\eea
The integral in the second line vanishes exactly and our expression for the superfluid density reduces to
\bea 
\rho_s &=&2\sum_{\bk}\frac{\Delta^2_0}{E^2_{\bk}}\left[\frac{1-2f}{2E_{\bk}} + \frac{\partial f}{\partial E_{\bk}}\right]k^2_z.\label{rhosmf3}\eea

Rearranging the mean-field expression for the mass density $\rho$ in Eq.~(\ref{rhomf}) using integration by parts and Eq.~(\ref{dervid}) (see the related discussion in Ref.~\onlinecite{PhysRep}), $\rho$ can be written as
\bea \rho &=&-m_F\sum_{\bk}k_z\frac{\partial}{\partial k_z}\left[1-\frac{\xi_{\bk}}{E_{\bk}}(1-2f)\right]\nonumber\\
&=&2\sum_{\bk}\frac{\Delta^2_0}{E^2_{\bk}}\frac{1-2f}{2E_{\bk}}k^2_z-2\sum_{\bk}\frac{\xi^2_{\bk}}{E^2_{\bk}}\frac{\partial f}{\partial E_{\bk}}k^2_z.\eea  Combining this result with Eq.~(\ref{rhosmf3}), we see that it reduces to Eq.~(\ref{Lrhos}).  Thus, evaluating Josephson's relation in Eq.~(\ref{JF}) using the mean-field approximation given by Eqs.~(\ref{F11mf}) and (\ref{F12mf}) gives us Landau's formula for the superfluid density in a BCS superfluid.

In Sec.~\ref{secBEC}, we employ the same mean-field approximation used in this section to show that the Josephson relation for a Fermi superfluid reduces to the analogous expression given by Eq.~(\ref{Josephson}) for a Bose superfluid, in the BEC limit of the BCS-BEC crossover.  

\section{Josephson's relation in the BEC limit} 
\label{secBEC}
An obvious feature of the Josephson relation for a Fermi superfluid is that it must reduce to Eq.~(\ref{Josephson}) in the BEC limit of the BCS-BEC crossover, where the Cooper pairs are tightly-bound dimer molecules.  
In this limit, where the $s$-wave scattering length $a_s$ is small and positive, the chemical potential becomes large and negative, roughly equal to half the dimer binding energy:~\cite{Engelbrecht} $\mu = -1/2m_Fa^2_s$.  In this case, $|\mu| \gg \Delta_0, k_BT$; $f \rightarrow 0$, and 
Eq.~(\ref{nc}) can be solved analytically to give the condensate density of dimer molecules,~\cite{Pieri03,Salasnich}
\bea n_c(T) \simeq \sum_{\bk}\frac{\Delta^2_0(T)}{4\xi^2_{\bk}}
\simeq \left(\frac{m^2_F a_s}{8\pi}\right)\Delta^2_0(T).\label{ncBEC}\eea
Within the same mean-field approximation [given by Eqs.~(\ref{F11mf}) and (\ref{F12mf})], one can show in the BEC limit that the inverse pair fluctuation propagator $\bL^{-1}$ reduces to~\cite{Pieri03,PieriPopov}
\bea \bL^{-1}(\bq,\nu_m) &=& \left(\frac{m^2_F a_s}{8\pi}\right)\mathbf{D}^{-1}(\bq,\nu_m),\label{GBEC}\eea
where  
\bea \lefteqn{\mathbf{D}^{-1}(\bq,\nu_m)\equiv}&&\nonumber\\&&
\!\!\!\!\left (
\begin{array}{cc} i\nu_m - \varepsilon_{\bq} -n_cU_{\mathrm{mol}} &
-n_cU_{\mathrm{mol}} 
\\
-n_cU_{\mathrm{mol}} 
&-i\nu_m - \varepsilon_{\bq} - n_cU_{\mathrm{mol}} 
\end{array} \right) \label{Dtilde}\eea is the inverse single-particle Green's function for the Bose-condensed dimer molecules, analogous to the Green's function that enters Eq.~(\ref{Josephson}).  Here, $\varepsilon_{\bq} = \bq^2/2m_B$ while $U_{\mathrm{mol}} = 4\pi(2a_s)/m_B$ is the mean-field interaction between dimers, which predicts a dimer scattering length of $a_{\mathrm{mol}} = 2a_s$~\cite{Engelbrecht} instead of the exact result $a_{\mathrm{mol}}=0.6a_s$.~\cite{PetrovMol}  Substituting Eqs.~(\ref{ncBEC}) and (\ref{GBEC}) into Eq.~(\ref{JF}), one immediately obtains the Josephson relation in Eq.~(\ref{Josephson}) for a condensate of dimer molecules.

Of course, one expects Eq.~(\ref{JF}) to reduce to Eq.~(\ref{Josephson}) in the BEC limit at any level of approximation, and not just at the mean-field level at which Eqs.~(\ref{GBEC}) and (\ref{Dtilde}) have been derived.  The fact that Eq.~(\ref{GBEC}) is a mean-field result only means that the molecular self-energies $\Sigma^{\mathrm{mol}}_{ij}(\bq,\nu_m)$ that enter the dimer Green's function $\mathbf{D}^{-1}$ in Eq.~(\ref{Dtilde}) are mean-field.   Explicitly, writing down the exact single-particle Green's function for a dimer molecule,
\bea \lefteqn{\mathbf{D}^{-1}(\bq,\nu_m)\equiv}&&\nonumber\\&&
\!\!\!\!\!\!\!\!\left (
\begin{array}{cc} i\nu_m- \varepsilon_{\bq}+ \mu_{\mathrm{mol}}-\Sigma^{\mathrm{mol}}_{11}&
-\Sigma^{\mathrm{mol}}_{12}
\\
-\Sigma^{\mathrm{mol}}_{21}
&-i\nu_m-\varepsilon_{\bq} +\mu_{\mathrm{mol}}-\Sigma^{\mathrm{mol}}_{22}
\end{array} \right),\nonumber\\ \label{D}\eea 
Eq.~(\ref{Dtilde}) corresponds to the result
\bea \mu_{\mathrm{mol}}-\Sigma^{\mathrm{mol}}_{11}(\bq,\nu_m) = -n_cU_{\mathrm{mol}}\label{selfenergy1}\eea
and
\bea \Sigma^{\mathrm{mol}}_{12}(\bq,\nu_m) = n_cU_{\mathrm{mol}}.\label{selfenergy2}\eea

We anticipate that, going past the mean-field approximation used to obtain the results in Eqs.~(\ref{selfenergy1}) and (\ref{selfenergy2}), one still arrives at the identity given by Eq.~(\ref{GBEC}), except that the self-energies will incorporate beyond-mean-field contributions.  Using Eq.~(\ref{D}) in Eq.~(\ref{Josephson}) shows that a momentum-independent self-energy will always lead to the well-known mean-field result, $\rho_s = m_Bn_c(T)$.  As first shown in Ref.~\onlinecite{Pieri03}, including contributions from fluctuations, the superfluid density in the BEC limit is actually given by Landau's formula for a normal fluid comprised of gapless Bogoliubov excitations of the BEC of dimer molecules.

\section{Summary}

Josephson's relation for Bose superfluids gave the first explicit identity connecting the two key order parameters in the theory of superfluids: the broken-symmetry order parameter and the superfluid density.  It is remarkable that such a simple relation exists between two such different quantities: the superfluid density that describes the response of a system to a transverse current probe, as in Eq.~(\ref{JJ}), and the order-parameter in a Bose superfluid, associated with the macroscopic occupation of a single-particle state.  Extending the recent analysis by Holzmann and Baym,~\cite{Holzmann} the analogous identity has been derived for a two-component $s$-wave Fermi superfluid. This gives an exact relation between the superfluid density $\rho_s$ and the BCS order parameter $\Delta_0$ in terms of the infrared limit of the static pair fluctuation propagator $\bL(\bq,\nu_m=0)$.  

Using mean-field BCS theory to evaluate the pair fluctuation propagator, we have seen that the Josephson relation derived in this paper reduces to the Josephson relation for a Bose superfluid in the BEC limit and also Landau's formula for the superfluid density of a Fermi gas with BCS quasiparticle excitations.  At first glance, it might seem surprising that the Josephson relation--which expresses the superfluid density in terms of the propagator for \textit{collective phase fluctuations}--manages to reproduce this Landau formula for a normal fluid of \textit{single-particle} Fermi BCS excitations.  However, this propagator is actually a correlation function for the \textit{gradient} of the phase of the order parameter,~\cite{note5} and consequently, is directly related to the current correlation function.~\cite{Baymbook}  In turn, it is well-known that Landau's formula can be obtained by a direct evaluation of the longitudinal and transverse components of the current correlation function within the BCS approximation.~\cite{FW}

The simple structure of the Josephson relation derived in this paper should simplify the calculation of the superfluid density using the standard tools of diagrammatic perturbation theory developed for the BCS-BEC crossover problem~\cite{Pieri04,HL,He} to evaluate the pair fluctuation propagator $\bL$.  It also opens the way to giving a rigorous analysis of the superfluid transition in Fermi systems, along the lines of those given for Bose superfluids.~\cite{Josephson,Holzmann03,Holzmann05}

\begin{acknowledgments}
I would like to thank A.~Griffin and L.~P.~Pitaevskii for illuminating discussions, and also M. Holzmann for clarifying some details of Ref.~\onlinecite{Holzmann}. 

\end{acknowledgments}


\begin{thebibliography}{99}
\bibitem{note4} The pair fluctuation propagator can be transformed via unitary transformation into a $2\times2$ matrix propagator describing fluctuations of the phase and amplitude of the order parameter.  In the long wavelength limit, $(\bL^{-1})_{11} - (\bL^{-1})_{12}$ is the inverse propagator for phase fluctuations.~\cite{Engelbrecht}
\end{thebibliography}
\end{document}